\documentclass[aps,preprint,prd,showpacs,nofootinbib]{revtex4}
\usepackage{amsmath}
\usepackage{graphicx}
\usepackage{dcolumn}
\usepackage{bm}
\usepackage{amssymb}
\usepackage{latexsym}

\def\be{\begin{equation}}
\def\ee{\end{equation}}
\def\ba{\begin{eqnarray}}
\def\ea{\end{eqnarray}}

\begin{document}

\title{ Thermal Tachyon Cosmology
}

\author{Yun-Song Piao$^{a,c}$ and Yuan-Zhong Zhang$^{b,c}$}
\affiliation{${}^a$College of Physical Sciences, Graduate School
of Chinese Academy of Sciences, YuQuan Road 19{\rm A}, Beijing
100049, China} \affiliation{$^b$CCAST (World Lab.), P.O. Box 8730,
Beijing 100080}\affiliation{$^{c}$Institute of Theoretical
Physics, Chinese Academy of Sciences, P.O. Box 2735, Beijing
100080, China }

\begin{abstract}
We show that in a multi
 D\={D} branes system with high temperature, there may exist a thermal
cosmological phase before usual tachyon inflation. Though this
thermal phase can be very transitory, it may has some interesting
applications for early tachyon/brane cosmology.

\end{abstract}

 \maketitle

Recently many efforts have been devoted to studying non-BPS
configurations of branes, such as D\={D} brane pairs, non-BPS
branes \cite{S, SZ, WS, SS}. For such a configuration, the
spectrum of open strings will contain a tachyon field, for non-BPS
brane it is real and for D\={D} brane pairs it is complex, which
indicates that this configuration is unstable. The tachyon field
will roll down from the vacuum of open string toward the minimum
of the tachyon effective potential. There has been a lot of
studies on various cosmological effects of the rolling tachyon
\cite{GMP, KL, FC, PC, MCI}.

However, for generic initial conditions, an early universe with
more high temperature may be more possible. Thus it is interesting
to investigate the cosmology of this kind of non-BPS configuration
in high temperature. It has been shown that in high temperature
the non-BPS configuration can become stable and not decay
\cite{DGK, WH, H}. Thus in analogy with high temperature symmetry
restoration in usual field theories \cite{DJ}, see \cite{LS} for a
thermal inflation, there may be a thermal cosmological evolutive
phase before the usual tachyon inflation, especially when the
number of branes is very large. The interest of this thermal
evolution lies in providing an initial condition for subsequent
tachyon inflation. We will show and discuss it in the following.


We focus on a system of N D3-\={D}3 branes pairs with finite
temperature. The free energy for the N D3-\={D}3 branes and mixed
gas system at temperature $T= \beta^{-1}$ can be written as,
following Ref. \cite{DGK},  \ba f(\phi, \beta) & = & 2
\tau_3\left( {\rm Tr}~ e^{-2|t|^2}\right) \nonumber\\
& + & {c N^2 \Omega\over (2\pi)^3 \beta^4} \int_0^\infty x^2
\ln\left(1-e^{-\sqrt{x^2+\beta^2 m^2}}\right) dx, \label{fphi} \ea
where \be \tau_3 = {M_s^4\over (2\pi)^3 g_s}\label{tau}\ee is D3
or \={D}3 brane tension, $t$ is related to the background tachyon
field $\phi$ appearing in the D brane worldsheet action through an
error function as follows \be |\phi| =\sqrt{\pi\over 2}{\rm
Erf}(|t|), \ee and $\Omega$ is the volume of a unit two-sphere,
the constant $c$ is given by $c= 8+ 8(7/8)$ for the relevant eight
bosonic and eight fermionic degrees of freedom, which include
gauge fields, transverse scalars and superpartners and form a
mixed gas system, $m\sim M_s |\phi|$ is the mass given to gauge
fields by Higgs effect. If the tachyon fields start at $\phi =0$,
then the tachyon rolling along the effective potential will lower
the energy of the system, but it also gives mass to the relative
gauge fields and decrease the entropy of the mixed gas system.
Thus, an amount $\delta \phi$ gives mass to $N$ out of the $N^2$
species of particles in the gas. Varying (\ref{fphi}), one have
\be \delta f(\phi, \beta) = -2\tau_3 (\delta \phi)^2 +{15\over 24}
N \beta^{-2}(\delta m)^2 \ee Therefore, for large enough
temperature \be T\geq T_c ={T_h\over \sqrt{g_s N}} \label{th} \ee
where the numerical constants have been disregarded, and $T_c$ is
regarded as a critical temperature, $\delta f> 0$, {\it i.e.} the
open string vacuum becomes a minimum of the free energy and is
stable. In this case, the tachyon field is no longer tachyonic and
the D\={D} brane pairs do not annihilate. $T_h\sim M_s$ is the
Hagedorn temperature, for open string, which is a limiting
temperature, {\it i.e.} in all senses $ T< T_h $. When approaching
this temperature, the energy is converted to the massive modes of
open string and the creation of D \={D} brane pairs may become
more important \cite{ABK}. Since the existence of the limiting
temperature, from (\ref{th}), we see that in usual case, $T$ can
hardly exceed $T_c$, only when $g_s N\gg 1$, $T> T_c$ becomes
possible. But in this case, open strings are strongly coupled,
which leads that the perturbative calculation used here becomes
distrusted. However we may take $g_sN\sim {\cal O}(1)$ marginally
for our purpose. In this case $T_c\sim T_h$. Thus when the
temperature $T\sim T_h$, the system will be stable. Further to
suppress closed string loops, $g_s\ll 1$ must be taken, which
makes the radiation process of closed string disfavored, and thus
the temperature of system is mainly relevant to open string. Thus
a large enough $N$ will be generally required, which may be also
required by some other cosmological consideration \cite{PC, MCI,
B, MC}.

Now we assume that $N\gg 1$ and $T_c\sim T_h$ and focus on the
cosmology of brane system consisting of $N$ D\=D brane pairs. In
this case, the open string vacuum is stable.
The energy density of mixed gas system considered is given by
\be \rho_{gas} \sim N^2 T^4 ,\ee since the number of massless mode
of open strings on $N$ D\={D} branes pairs is proportional to
$N^2$. Thus initially for $T\sim T_c\sim M_s$, we have
$\rho_{gas}\sim N^2 M_s^4$, in the meantime the energy density of
branes tensor is $\rho_{brane}\sim N\tau_3$. Thus considering
(\ref{tau}), we have \be \rho_{brane}\sim {NM_s^4\over (2\pi)^3
g_s} \sim {N^2 M_s^4\over (2\pi)^3}\sim 10^{-2} \rho_{gas} , \ee
where $g_sN \sim {\cal O}(1)$ has been used. This result indicates
that when the temperature of the brane system approaches the
Hagedorn temperature $T_h$, the energy density of mixed gas system
will be dominated, while the branes tension is subdominated.
Thus generally there will be a radiation-dominated thermal phase
in the tachyon/brane cosmology before the tachyon becomes
tachyonic. During this period, \be H^2 \sim {N^2 M_s^4\over
M_p^2}={g_s^2 N^2 \over v} M_s^2, \ee where in second equation, we
use $M_p^2 ={v M_s^2\over g_s^2}$, where $v=(M_s R)^6\gg 1$ is the
volume of extra spaces in string unit. Thus for $g_sN \sim {\cal
O}(1)$, we have ${1\over H}\gg l_s$, where $l_s= {1\over M_s}$ is
the string length. Therefore, during the cosmological evolution of
brane system, the description of 4D effective field theory for
tachyon/brane cosmology seems reasonable.

In the following we will use the 4D effective description for the
cosmology of above brane system. Initially the temperature $T\sim
T_c\sim T_h$, and the brane cosmology is dominated by the mixed
gas system with the energy density $\rho\sim 1/a^4$ where $a$ is
the scale factor of expanding brane cosmology. In this case the
temperature $T\propto {1\over a}$, and thus will decrease rapidly
with the cosmological expansion. When the temperature drops to
$T\sim 0.3 T_h$, in which $\rho_{gas}\simeq \rho_{brane}$, the
energy density of branes tensor will begin to dominated the
universe. This means that \be H^2 \simeq {2N\tau_3 \over M_p^2},
\ee and is a nearly constant. Thus the universe will enter into an
inflation stage, as usual tachyon inflation. In this stage, since
$T\leq T_h\sim T_c$ the open string vacuum has became unstable and
will evolve toward closed vacuum, {\it i.e.} the minimum of
tachyon potential. The relevant cosmology with large $N$ branes
has been discussed in Ref. \cite{PC, MCI}.
Note also that during the radiation-dominated, $\rho\sim 1/t^2$.
Thus the time of existence of this thermal phase is given by \be
t\sim (T^2_h/T^2) t_c \sim 10 t_c, \ee which seems very
transitory. However, the transitory appearance of this thermal
evolutive phase before usual tachyon inflation may be interesting
for some purposes as follows,
Imagining a model where branes warp some cycles of the compact
manifold, and the volume of cycles is generally not constant due
to the pinching singularities of compact manifold, we can see that
branes and antibranes will minimize their energy and move towards
points in which the cycles have minimal volume \cite{EGJ}. The
kind of singularities or points can be regarded as collectors of
branes and antibranes, in which branes will be collected and
coincident. The occurrence of thermal expanding phase will be
helpful to remove singularities or points of small cycle volume in
generic compact manifolds and prevent early collapse of brane
universe, In addition, since in $T\sim T_h\sim T_c$ the open
string vacuum is a minimum and is stable again decay, it is easily
localized $\phi$ into $\phi =0$ from the viewpoint of effective
field theory. This in some sense corresponds to provide an initial
conditions for subsequent tachyon inflation.

In summary, we discuss the cosmological evolution of a multi
D\={D} branes system with high temperature, and its possible
application for early universe.
This work implies that the brane cosmology with high temperature
may have some interesting phenomena, which is worth studying
further.

\textbf{Acknowledgments} The author would like to thank Qing-Guo
Huang and Miao Li for earlier helpful discussions. This work is
supported in part by NNSFC under Grant Nos: 10405029 and 90403032,
as well as in part by National Basic Research Program of China
under Grant No: 2003CB716300.

\end{document}